\documentclass[aps,prl,twocolumn,superscriptaddress]{revtex4}
\usepackage{amssymb,amsbsy,graphicx,xcolor,times,subfigure,marvosym,color,bm,multirow,epstopdf}
\usepackage[latin1]{inputenc}
\begin{document}

\title{Resonant inelastic X-ray scattering response of the Kitaev honeycomb model}

\author{G\'abor B. Hal\'asz}
\affiliation{Kavli Institute for Theoretical Physics, University of
California, Santa Barbara, CA 93106, USA}

\author{Natalia B. Perkins}
\affiliation{School of Physics and Astronomy, University of
Minnesota, Minneapolis, MN 55116, USA}

\author{Jeroen van den Brink}
\affiliation{IFW Dresden, Helmholtzstrasse 20, 01069 Dresden,
Germany} \affiliation{Department of Physics, Harvard University,
Cambridge, MA 02138, USA}

%%%%%%%%%%%%%%%%%%%%%%%%%%%%%%%%%%%%%%%%%%%%%%%%%

\begin{abstract}

We calculate the resonant inelastic X-ray scattering (RIXS) response
of the Kitaev honeycomb model, an exactly solvable
quantum-spin-liquid model with fractionalized Majorana and flux
excitations. We find that the fundamental RIXS channels, the
spin-conserving (SC) and the non-spin-conserving (NSC) ones, do not
interfere and give completely different responses. SC-RIXS picks up
exclusively the Majorana sector with a pronounced momentum
dispersion, whereas NSC-RIXS also creates immobile fluxes, thereby
rendering the response only weakly momentum dependent, as in the
spin structure factor measured by inelastic neutron scattering. RIXS
can therefore pick up the fractionalized excitations of the Kitaev
spin liquid separately, making it a sensitive probe to detect
spin-liquid character in potential material incarnations of the
Kitaev honeycomb model.

\end{abstract}

%%%%%%%%%%%%%%%%%%%%%%%%%%%%%%%%%%%%%%%%%%%%%%%%%

\maketitle

%%%%%%%%%%%%%%%%%%%%%%%%%%%%%%%%%%%%%%%%%%%%%%%%%

Quantum spins in a solid can, instead of ordering in a definite
pattern, form a fluid type of ground state: a quantum spin liquid
(QSL) \cite{QSL}. Theory predicts a remarkable set of collective
phenomena to occur in such QSLs, including topological ground-state
degeneracy, long-range entanglement, and fractionalized excitations.
Beyond their clear theoretical appeal, these exotic properties also
find applications in the field of topological quantum computing
\cite{TQC}.

\begin{figure}[h]
\centering
\includegraphics[width=0.95\columnwidth]{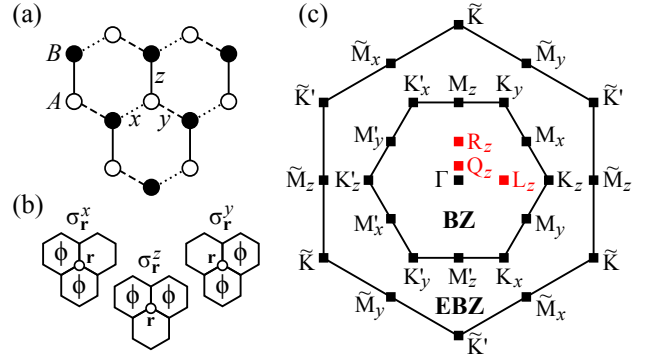}
\caption{(a) Illustration of the honeycomb lattice. Sites in
sublattice $A$ ($B$) are marked by white (black) circles, while $x$,
$y$, and $z$ bonds are marked by dotted, dashed, and solid lines,
respectively. (b) Flux excitations $\phi$ around the
photon-scattering site $\mathbf{r}$ (white circle) in the final
states $| m \rangle$ of the three fundamental NSC-RIXS channels with
amplitudes $\propto \langle m | \sigma_{\mathbf{r}}^{x,y,z} e^{-i t
\tilde{H} (\mathbf{r})} | 0 \rangle$ [see Eq.~(\ref{eq-A-2})],
respectively. (c) Illustration of the standard Brillouin zone (BZ),
and the extended Brillouin zone (EBZ) with respect to which the RIXS
response is periodic. The SC-RIXS response in Fig.~\ref{fig-3} is
plotted at special points (black squares) \cite{EBZ} and generic
representative points (red squares). \\} \label{fig-1}
\end{figure}

The Kitaev honeycomb model is an exactly solvable yet realistic spin
model with a QSL ground state \cite{Kitaev}. Neighboring $S = 1/2$
spins $\sigma_{\mathbf{r}}^{x,y,z}$ at the sites $\mathbf{r}$ of the
honeycomb lattice are coupled via different spin components along
the three bonds connected to any given site. The Hamiltonian is then
\begin{equation}
H_K = -J_x \sum_{\langle \mathbf{r}, \mathbf{r}' \rangle_x}
\sigma_{\mathbf{r}}^x \sigma_{\mathbf{r}'}^x - J_y \sum_{\langle
\mathbf{r}, \mathbf{r}' \rangle_y} \sigma_{\mathbf{r}}^y
\sigma_{\mathbf{r}'}^y - J_z \sum_{\langle \mathbf{r}, \mathbf{r}'
\rangle_z} \sigma_{\mathbf{r}}^z \sigma_{\mathbf{r}'}^z,
\label{eq-H}
\end{equation}
where $J_{x,y,z}$ are the coupling constants for the three types of
bonds $x$, $y$, and $z$ [see Fig.~\ref{fig-1}(a)]. Depending on
$J_{x,y,z}$, the model has two distinct phases. In the gapped
(gapless) phase, the ground state is a gapped (gapless) QSL. The
spins fractionalize into two types of elementary excitations in both
phases: Majorana fermions and emergent gauge fluxes.

From an experimental standpoint, finding a physical realization of
the Kitaev honeycomb model has proven to be a challenging task. So
far, three types of honeycomb systems have been proposed as
candidate incarnations of $H_K$: the iridates $\alpha$-A$_2$IrO$_3$
with A $=$ Na or Li \cite{Jackeli, Chaloupka, Ir-exp}, the ruthenate
$\alpha$-RuCl$_3$ \cite{Ru-exp, Banerjee}, and ultracold atoms in
optical lattices \cite{Duan}. In the iridates and the ruthenate,
however, a potential spin-liquid phase at low temperatures is
preempted by magnetic order due to residual magnetic interactions
beyond $H_K$ \cite{Chaloupka, Extra}. Nevertheless, since these
interactions are typically small, the higher-energy excitations
above the energy scale setting the magnetic order are expected to be
governed by $H_K$ \cite{Banerjee}.

Given that QSLs are inherently defined in terms of a property that
they do not have (i.e., magnetic order), their experimental
identification and characterization are far from obvious \cite{QSL}.
One potential hallmark of QSLs is the presence of fractionalized
magnetic excitations. For the Kitaev spin liquid, it has been
proposed that signatures of these excitations can be observed by
inelastic neutron scattering (INS) \cite{Baskaran, Knolle-1} and by
Raman scattering (RS) with visible light \cite{Knolle-2}. However,
both of these methods have important limitations. In particular,
neither of them can directly probe the highly dispersive gapless
Majorana excitations. INS displays an overall energy gap and shows
little momentum dispersion because it creates two immobile flux
excitations that dominate the response. RS creates two Majorana
excitations only and measures their density of states, but it is an
inherently zero-momentum probe and does not provide any information
on their dispersion.

Using the exact solution of the Kitaev honeycomb model, we
demonstrate in this Letter that resonant inelastic X-ray scattering
(RIXS) can probe each type of fractionalized excitation directly and
independently. We establish that the four fundamental RIXS channels,
the spin-conserving (SC) and the three non-spin-conserving (NSC)
ones, do not interfere and give completely different responses. The
SC-RIXS channel does not create any fluxes and picks up exclusively
the Majorana fermions with a pronounced momentum dispersion.
Conversely, the NSC-RIXS channels involve flux creation and
therefore show little momentum dependence. In the physical regime,
they are found to map onto the respective components of the spin
structure factor measured by INS. Since the RIXS response directly
quantifies both Majorana and flux excitations, it can serve as an
effective probe of Kitaev-spin-liquid character in any experimental
candidate material.

\emph{Formalism.---}When calculating the RIXS response, we consider
the $L_3$-edge of the $\alpha$-A$_2$IrO$_3$ iridates with the
Ir$^{4+}$ ion being in a $5d^5$ configuration. However, as we later
argue, our results directly translate to other edges of Ir$^{4+}$ or
Ru$^{3+}$ and also to similar responses in ultracold atomic systems.
RIXS is a second-order process consisting of two dipole transitions
\cite{RIXS}. First, a photon is absorbed, and an electron from the
$2p$ core shell is excited into the $5d$ valence shell, thereby
creating a $2p$ core hole and an extra $5d$ electron. Second, an
electron from the $5d$ valence shell decays into the $2p$ core hole,
and a photon is emitted. The low-energy physics of the $5d$
electrons at each Ir$^{4+}$ ion is governed by a $J = 1/2$ Kramers
doublet in the $t_{2g}$ orbitals, and we assume that $H \equiv H_K$
is the effective low-energy Hamiltonian acting on these Kramers
doublets \cite{Jackeli}. In terms of the corresponding Kitaev model,
the $5d^6$ configuration in the intermediate state is then described
as a vacancy or, equivalently, a non-magnetic impurity.

The initial and the final states of the RIXS process are $| 0
\rangle \otimes | \mathbf{Q}, \bm{\epsilon} \rangle$ and $| m
\rangle \otimes | \mathbf{Q}', \bm{\epsilon}' \rangle$,
respectively, where $| 0 \rangle$ is the ground state of the Kitaev
model, $| m \rangle$ is a generic eigenstate with energy $E_m$ with
respect to $| 0 \rangle$, while $\mathbf{Q}$ ($\mathbf{Q}'$) is the
momentum and $\bm{\epsilon}$ ($\bm{\epsilon}'$) is the polarization
of the incident (scattered) photon. During the RIXS process, a
momentum $\mathbf{q} \equiv \mathbf{Q} - \mathbf{Q}'$ and an energy
$\omega = E_m$ is transferred from the photon to the Kitaev spin
liquid. The total RIXS intensity is $I (\omega, \mathbf{q}) = \sum_m
| \sum_{\alpha, \beta} T_{\alpha \beta} A_{\alpha \beta} (m,
\mathbf{q})|^2 \, \delta (\omega - E_m)$, where $T_{\alpha \beta}$
is a spin-space polarization tensor depending on the microscopic
details of the RIXS process (i.e., the ion type, the edge type, and
the photon polarizations), and $A_{\alpha \beta} (m, \mathbf{q})$ is
the scattering amplitude from $| 0 \rangle \otimes | \mathbf{Q},
\bm{\epsilon} \rangle$ to $| m \rangle \otimes | \mathbf{Q}',
\bm{\epsilon}' \rangle$. This amplitude is given by the
Kramers-Heisenberg formula:
\begin{equation}
A_{\alpha \beta} (m, \mathbf{q}) = \sum_{\mathbf{r},
\tilde{n}_{\mathbf{r}}} \frac{\langle m | d_{\mathbf{r},
\alpha}^{\phantom{\dag}} | \tilde{n}_{\mathbf{r}} \rangle \langle
\tilde{n}_{\mathbf{r}} | d_{\mathbf{r}, \beta}^{\dag} | 0 \rangle}
{\Omega - E_{\tilde{n}} + i \Gamma} \, e^{i \mathbf{q} \cdot
\mathbf{r}}, \label{eq-A-1}
\end{equation}
where $\Gamma$ is the inverse lifetime of the core hole, $\Omega$ is
the energy of the incident photon with respect to the resonance
energy (i.e., the energy difference between the $5d$ and the $2p$
shells), and the operator $d_{{\mathbf{r}}, \sigma}^{\dag}$ promotes
a $2p$ electron with spin $\sigma$ at site ${\mathbf{r}}$ into a
$5d$ state at the same site. In terms of the Kitaev model, this
operation is equivalent to an electron with spin $-\sigma$ being
annihilated at site ${\mathbf{r}}$. The intermediate state
$|\tilde{n}_{\mathbf{r}} \rangle$ is then a generic eigenstate of
the Kitaev model with a single vacancy at site ${\mathbf{r}}$ that
has energy $E_{\tilde{n}}$ with respect to the ground state $|
\tilde{0}_{\mathbf{r}} \rangle$ of the same model. Note that we use
a tilde to distinguish the model with a vacancy (intermediate
states) from the one without a vacancy (initial and final states).

The four fundamental RIXS channels are introduced by decomposing the
polarization tensor into $T_{\alpha \beta} = P_{\eta} \sigma_{\alpha
\beta}^{\eta}$ with $\eta = \{ 0,x,y,z \}$, where $\sigma^0$ is the
identity matrix, and $\sigma^{x,y,z}$ are the Pauli matrices. In the
spin-conserving (SC) channel with $T_{\alpha \beta} \propto
\sigma_{\alpha \beta}^0$, the spin of the $5d$ valence shell does
not change during the RIXS process, while in the three
non-spin-conserving (NSC) channels with $T_{\alpha \beta} \propto
\sigma_{\alpha \beta}^{x,y,z}$, the same spin is rotated by $\pi$
around the $x,y,z$ axes, respectively. For the $L_3$-edge of the
Ir$^{4+}$ ion, the SC coefficient is $P_0 = \bm{\epsilon}'^{*} \cdot
\bm{\epsilon}$, while the NSC coefficients are $P_x = i
(\epsilon'^{*}_y \epsilon_z - \epsilon'^{*}_z \epsilon_y)$ and
$P_{y,z}$ its cyclic permutations \cite{Ament}.

Our first main result is that the four RIXS channels do not
interfere in the case of the Kitaev model because they result in
mutually orthogonal final states. In particular, the final state has
no flux excitations for the SC-RIXS channel, while it has two flux
excitations separated by $x,y,z$ bonds for the three NSC-RIXS
channels, respectively [see Fig.~\ref{fig-1}(b)]. We provide a
detailed derivation of this result in the Supplementary Material
(SM). For any scattering geometry and polarizations, the total RIXS
intensity $I (\omega, \mathbf{q})$ is then a sum of four individual
intensities $I_{\eta} (\omega, \mathbf{q}) = \sum_m |A_{\eta} (m,
\mathbf{q})|^2 \, \delta (\omega - E_m)$ corresponding to the four
channels $\eta = \{ 0,x,y,z \}$. It is derived in the SM that the
individual RIXS amplitudes are
\begin{equation}
A_{\eta} (m, \mathbf{q}) \propto \sum_{\mathbf{r}} \int_0^{\infty}
dt \, e^{-\Gamma t + i \Omega t + i \mathbf{q} \cdot \mathbf{r}}
\langle m | \sigma_{\mathbf{r}}^{\eta} e^{-i t \tilde{H}
(\mathbf{r})} | 0 \rangle, \label{eq-A-2}
\end{equation}
where $\tilde{H} (\mathbf{r}) = H + \sum_{\kappa = x,y,z} J_{\kappa}
\sigma_{\mathbf{r}}^{\kappa} \sigma_{\kappa (\mathbf{r})}^{\kappa}$
is the Hamiltonian of the Kitaev model with a single vacancy at site
$\mathbf{r}$. The spin at site $\mathbf{r}$ is effectively removed
from the model by being decoupled from its neighbors at sites
$\kappa (\mathbf{r})$ \cite{Halasz-1}.

Since the inverse lifetime $\Gamma$ is by far the largest energy
scale in both the iridates $\alpha$-A$_2$IrO$_3$ \cite{Ir-scale} and
the ruthenate $\alpha$-RuCl$_3$ \cite{Banerjee, Ru-scale}, we employ
the fast-collision approximation to RIXS, for which $\Gamma
\rightarrow \infty$ and hence $t \sim 1 / \Gamma \rightarrow 0$.
Expanding $e^{-i t \tilde{H} (\mathbf{r})}$ up to first order in
$J_{x,y,z} / \Gamma$, integrating over $t$, and demanding $H | 0
\rangle = 0$ by adding a trivial constant term to $H$, the RIXS
amplitudes in Eq.~(\ref{eq-A-2}) become
\begin{eqnarray}
A_{\eta} (m, \mathbf{q}) &\propto& \sum_{\mathbf{r}} e^{i \mathbf{q}
\cdot \mathbf{r}} \langle m | \sigma_{\mathbf{r}}^{\eta} \bigg[ 1 -
\frac{i \tilde{H} (\mathbf{r})} {\Gamma} \bigg] | 0
\rangle \label{eq-A-3} \\
&=& \sum_{\mathbf{r}} e^{i \mathbf{q} \cdot \mathbf{r}} \langle m |
\sigma_{\mathbf{r}}^{\eta} \bigg[ 1 - \frac{i}{\Gamma} \sum_{\kappa}
J_{\kappa} \sigma_{\mathbf{r}}^{\kappa} \sigma_{\kappa
(\mathbf{r})}^{\kappa} \bigg] | 0 \rangle, \nonumber
\end{eqnarray}
where we also set $\Omega = 0$ for simplicity by recognizing that
its exact value does not matter as long as $\Omega \ll \Gamma$. We
emphasize that the final form of Eq.~(\ref{eq-A-3}) is expected to
be generic beyond the $L_3$-edge of the Ir$^{4+}$ ion. For any
relevant RIXS process, the couplings in the intermediate state are
perturbed (i.e., weakened or switched off) around the
photon-scattering site $\mathbf{r}$, and an analogous calculation in
the fast-collision approximation would then give an identical
first-order result, up to a potential renormalization of $\Gamma$.

Given that the Kitaev model is exactly solvable both in the presence
and in the absence of a vacancy \cite{Halasz-1, Willans}, the RIXS
amplitudes in Eq.~(\ref{eq-A-3}) can be evaluated exactly. Our
calculation follows the usual procedure \cite{Kitaev}. We first take
care of the static (and local) fluxes, then introduce Majorana
fermions to obtain a quadratic fermion problem for each flux
configuration, and finally deal with the resulting free-fermion
problems by means of standard methods. However, due to technical
reasons (see SM), the RIXS intensities are calculated differently
for the SC and the NSC channels \cite{Knolle-3}.

For the SC channel, we evaluate $A_0 (m, \mathbf{q})$ for each
individual final state $| m \rangle$ and obtain $I_0 (\omega,
\mathbf{q})$ as a histogram of $|A_0 (m, \mathbf{q})|^2$ in terms of
the final-state energies $\omega = E_m$. In the language of
Ref.~\onlinecite{Knolle-3}, this method corresponds to the
few-particle approach. Indeed, it follows from Eq.~(\ref{eq-A-3})
that, up to first order in $J_{x,y,z} / \Gamma$, there are two types
of final states with a non-zero RIXS amplitude: $| m \rangle = | 0
\rangle$ with no excitations at all, and $| m \rangle \neq | 0
\rangle$ with no flux and two fermion excitations. Since scattering
back into the ground state $| 0 \rangle$ corresponds to a purely
elastic process, we restrict our attention to the $| m \rangle \neq
| 0 \rangle$ final states with two fermion excitations at momenta
$\mathbf{k}$ and $\mathbf{q} - \mathbf{k}$. The energy of such a
state is $E_m = \varepsilon_{\mathbf{k}} + \varepsilon_{\mathbf{q} -
\mathbf{k}}$, where $\varepsilon_{\mathbf{k}} = 2
|\lambda_{\mathbf{k}}|$ is the energy of a single fermion, and
$\lambda_{\mathbf{k}} \equiv \sum_{\kappa = x,y,z} J_{\kappa} e^{i
\mathbf{k} \cdot \hat{\mathbf{r}}_{\kappa}}$ in terms of the three
bond vectors $\hat{\mathbf{r}}_{x,y,z}$ pointing from any site in
sublattice $A$ to its respective neighbors in sublattice $B$ [see
Fig.~\ref{fig-1}(a)]. The SC-RIXS intensity is then derived in the
SM to be
\begin{eqnarray}
I_0 (\omega, \mathbf{q}) &\propto& \int_{\mathrm{BZ}} d^2 \mathbf{k}
\,\, \delta (\omega - \varepsilon_{\mathbf{k}} -
\varepsilon_{\mathbf{q} - \mathbf{k}}) \left[
\varepsilon_{\mathbf{k}} - \varepsilon_{\mathbf{q} - \mathbf{k}}
\right]^2 \nonumber \\
&& \times \left| 1 - e^{i \varphi_{\mathbf{k}}} \, e^{i
\varphi_{\mathbf{q} - \mathbf{k}}} \right|^2, \label{eq-I}
\end{eqnarray}
where $e^{i \varphi_{\mathbf{k}}} \equiv \lambda_{\mathbf{k}} /
|\lambda_{\mathbf{k}}|$ is a phase factor between the two
sublattices. Since the bond vectors $\hat{\mathbf{r}}_{x,y,z}$ are
not lattice vectors, the intensity is not periodic with respect to
the standard Brillouin zone (BZ), but with respect to the extended
Brillouin zone (EBZ) illustrated in Fig.~\ref{fig-1}(c).

\begin{figure}[b]
\includegraphics[width=\columnwidth]{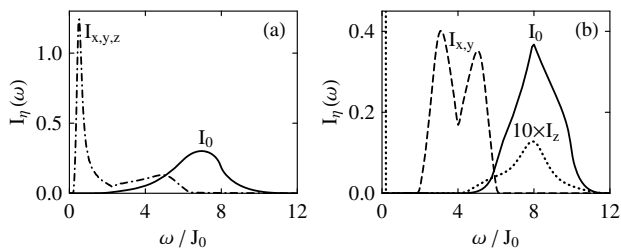}
\caption{Momentum-integrated RIXS intensities at (a) the isotropic
point of the gapless phase [$J_{x,y,z} = J_0$] and (b) a
representative point in the gapped phase [$J_{x,y} = J_0 / 2$ and
$J_z = 2 J_0$]. The SC response $I_0 (\omega)$ is plotted by a solid
line, while the NSC responses $I_{x,y,z} (\omega)$ are plotted by
dashed and/or dotted lines. Each response is normalized such that
$\int d\omega \, I_{\eta} (\omega) = 1$. At the gapped point, the
NSC response $I_z (\omega)$ has a delta peak at low energies and is
multiplied by $10$ to be comparable with the other responses. \\}
\label{fig-2}
\end{figure}

For the NSC channels, we consider the RIXS intensity directly and
rewrite it as $I_{\kappa} (\omega, \mathbf{q}) \propto
\int_{-\infty}^{+\infty} ds \, e^{i \omega s} K_{\kappa} (s,
\mathbf{q})$ in terms of a time-like variable $s$. It is derived in
the SM that the kernel of the resulting integral takes the form
\begin{eqnarray}
K_{\kappa} (s, \mathbf{q}) &=& \langle 0 | \left[
\sigma_{\mathbf{0}}^{\kappa} \, e^{i \tilde{H} (\mathbf{0}) /
\Gamma} + \sigma_{\hat{\mathbf{r}}_{\kappa}}^{\kappa} e^{i \tilde{H}
(\hat{\mathbf{r}}_{\kappa}) / \Gamma - i \mathbf{q} \cdot
\hat{\mathbf{r}}_{\kappa}} \right] e^{-i s H} \nonumber \\
&& \times \left[ \sigma_{\mathbf{0}}^{\kappa} \, e^{-i \tilde{H}
(\mathbf{0}) / \Gamma} + \sigma_{\hat{\mathbf{r}}_{\kappa}}^{\kappa}
e^{-i \tilde{H} (\hat{\mathbf{r}}_{\kappa}) / \Gamma + i \mathbf{q}
\cdot \hat{\mathbf{r}}_{\kappa}} \right] | 0 \rangle, \nonumber
\end{eqnarray}
where $\mathbf{0}$ is any site in sublattice $A$. In the language of
Ref.~\onlinecite{Knolle-3}, this method corresponds to the
determinant approach. Indeed, the ground-state expectation values in
$K_{\kappa} (s, \mathbf{q})$ can be evaluated as functional
determinants (see SM). We remark that the NSC-RIXS response
$I_{\kappa} (\omega, \mathbf{q})$ reduces to the corresponding
component $\kappa$ of the spin structure factor \cite{Knolle-1} in
the limit of $\Gamma \rightarrow \infty$. This result is in contrast
with SC-RIXS, where the inelastic response disappears in the same
limit.

\begin{figure*}[t]
\includegraphics[width=2.25\columnwidth]{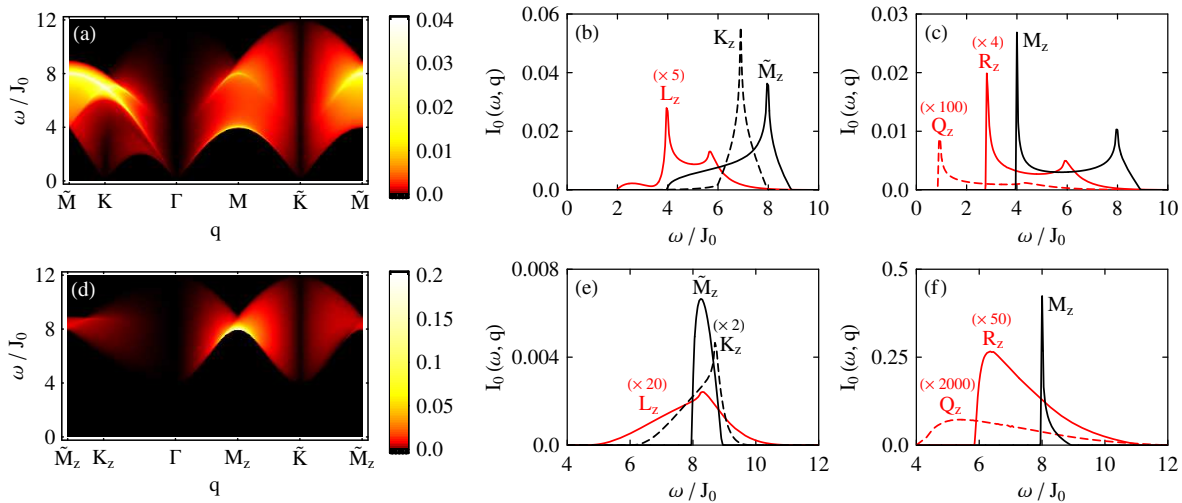}
\caption{Momentum-resolved SC-RIXS intensities at the isotropic
point of the gapless phase [$J_{x,y,z} = J_0$] (a, b, c) and at a
representative point in the gapped phase [$J_{x,y} = J_0 / 2$ and
$J_z = 2 J_0$] (d, e, f). The response is plotted along the entire
$\tilde{\mathrm{M}}_z$-$\mathrm{K}_z$-$\mathrm{\Gamma}$-$\mathrm{M}_z$-$\tilde{\mathrm{K}}$-$\tilde{\mathrm{M}}_z$
cut (a, d), at specific points on the
$\mathrm{\Gamma}$-$\mathrm{K}_z$-$\tilde{\mathrm{M}}_z$ cut (b, e),
and at specific points on the $\mathrm{\Gamma}$-$\mathrm{M}_z$ cut
(c, f) [see Fig.~\ref{fig-1}(c)]. The intensity is normalized such
that $\int d\omega \int d^2 \mathbf{q} \, I_0 (\omega, \mathbf{q}) =
1$, where $\mathbf{q}$ is measured in units of $a^{-1} \equiv
|\hat{\mathbf{r}}_{x,y,z}|^{-1}$ and is integrated over the entire
EBZ. Some responses are multiplied by numerical factors (written
next to them) to be comparable with other responses.} \label{fig-3}
\end{figure*}

\emph{Results.---}We first discuss the momentum-integrated RIXS
intensities $I_{\eta} (\omega) = \int_{\mathrm{EBZ}} d^2 \mathbf{q}
\, I_{\eta} (\omega, \mathbf{q})$. In Fig.~\ref{fig-2}, the SC and
NSC responses are plotted for representative points of both the
gapless (a) and the gapped (b) phases. All of our responses are
universal in the sense that their functional forms do not depend on
the precise value of $\Gamma \gg J_{x,y,z}$. Also, some or all of
the NSC responses can be identical due to symmetry. Since the
maximal fermion energy is $2 \sum_{\kappa} J_{\kappa} = 6 J_0$ at
both representative points, the responses with maximal energies
$\approx 6 J_0$ and $\approx 12 J_0$ can be identified as
predominantly one-fermion and two-fermion responses, respectively
\cite{Knolle-3}. Similarly, any delta peak close to zero energy
corresponds to a zero-fermion response. Unlike the NSC responses,
the SC response is always dominated by two-fermion excitations.
Furthermore, the SC and NSC responses have different low-energy
behavior in the gapless phase. The NSC response has an energy gap
due to flux creation, while the SC response is found to vanish as
$\propto \omega^5$ in the limit of $\omega \rightarrow 0$. Three
powers of $\omega$ come from the two-fermion density of states
around the Dirac points \cite{Song}, and two further powers appear
due to the factor $[\varepsilon_{\mathbf{k}} -
\varepsilon_{\mathbf{q} - \mathbf{k}}]^2$ in Eq.~(\ref{eq-I}), which
indicates that the fermions at lower energies are perturbed less by
the presence of the vacancy \cite{Halasz-2}.

The differences between the SC and NSC responses become even more
evident when we consider the momentum-resolved RIXS intensities
$I_{\eta} (\omega, \mathbf{q})$. In Fig.~\ref{fig-3}, the SC
response is plotted for the representative points of the two phases.
The fractionalized nature of the excitations is indicated by the
lack of delta peaks corresponding to well-defined $\omega
(\mathbf{q})$ dispersions in the spectrum. Nevertheless, the
response has a pronounced momentum dependence and is therefore able
to probe the dispersions of the individual excitations directly. For
example, in the gapless phase, the Dirac points $\mathrm{K}$ in the
fermion dispersion manifest themselves in gapless responses around
the $\mathrm{\Gamma}$, $\mathrm{K}$, and $\tilde{\mathrm{K}}$ points
of the EBZ. However, the response actually vanishes at the
$\mathrm{\Gamma}$ and $\tilde{\mathrm{K}}$ points due to the factor
$[\varepsilon_{\mathbf{k}} - \varepsilon_{\mathbf{q} -
\mathbf{k}}]^2$ in Eq.~(\ref{eq-I}) and the fermion dispersion being
symmetric around these points. There is a further depression of the
response around the $\mathrm{\Gamma}$ point due to a destructive
interference between the two sublattices, as indicated by the minus
sign in the factor $|1 - e^{i \varphi_{\mathbf{k}}} \, e^{i
\varphi_{\mathbf{q} - \mathbf{k}}}|^2$ of Eq.~(\ref{eq-I}). This
effect arises because the fermions transform projectively under
inversion and is therefore a direct signature of their
fractionalized nature \cite{You}. We also remark that, unlike the
NSC responses, the SC response is invariant under $J_{x,y,z}
\rightarrow -J_{x,y,z}$.

In contrast to the SC response, the NSC responses show little
momentum dependence because the localized fluxes created by them can
absorb momentum well. In fact, we find that the three NSC-RIXS
components are virtually indistinguishable from the corresponding
components of the spin structure factor \cite{Knolle-1} in the
$\Gamma / J_0 \gtrsim 100$ regime, which is physically relevant for
both $\alpha$-A$_2$IrO$_3$ \cite{Ir-scale} and $\alpha$-RuCl$_3$
\cite{Banerjee, Ru-scale}. NSC-RIXS can therefore fully determine
the spin structure factor in the iridates, for which INS is
challenging due to the large neutron-absorption cross section of
iridium. Although RIXS is currently limited by its energy resolution
$\Delta \omega \sim J_0$ \cite{RIXS-res}, this technique has been
improving rapidly, and therefore $\Delta \omega \ll J_0$ is a
distinct possibility for the near future.

We finally discuss the RIXS responses at a generic point of the
Kitaev-spin-liquid phase, which corresponds to a generic
time-reversal-invariant perturbation with respect to $H_K$. For the
NSC-RIXS channels, the results in Ref.~\onlinecite{Song} are
directly applicable and imply that the response is generically
gapless. For the SC-RIXS channel, a similar analysis indicates that
the response no longer vanishes at the $\mathrm{\Gamma}$ and
$\tilde{\mathrm{K}}$ points and that $I_0 (\omega)$ takes the
low-energy form of $\propto \omega^3$ instead of $\propto \omega^5$
in the most generic case. However, since the characteristic lower
edge of the spectrum in Fig.~\ref{fig-3} is robust, we expect that
SC-RIXS remains an effective probe of the fermion dispersion for a
generic Kitaev spin liquid. Furthermore, some higher-energy features
are believed to persist even beyond the phase transition into the
magnetically ordered phase \cite{Banerjee}.

\emph{Conclusions.---}Calculating the exact RIXS response of the
Kitaev honeycomb model, we have found that the four fundamental RIXS
channels, the SC and the three NSC ones, do not interfere and
correspond to completely different responses. In the physically
relevant regime, the SC response displays a pronounced momentum
dependence and picks up the gapless Majorana fermions, while the NSC
responses are only weakly momentum dependent and recover the
respective components of the spin structure factor. We therefore
believe that RIXS can serve as an effective probe of spin-liquid
character in present and future candidate materials for the
realization of the Kitaev honeycomb model.

We thank D.~A.~Abanin, J.~T.~Chalker, D.~V.~Efremov, and
I.~Rousochatzakis for useful discussions. G.~B.~H.~is supported by a
fellowship from the Gordon and Betty Moore Foundation (Grant
No.~4304). N.~P.~is supported by the NSF Grant No.~DMR-1511768.
J.~v.~d.~B.~acknowledges support from the Deutsche
Forschungsgemeinschaft via Grant No.~SFB 1143 and the Harvard-MIT
CUA. This research was supported in part by the National Science
Foundation under Grant No.~NSF PHY11-25915.

%%%%%%%%%%%%%%%%%%%%%%%%%%%%%%%%%%%%%%%%%%%%%%%%%

%%%%%%%%%%%%%%%%%%%%%%%%%%%%%%%%%%%%%%%%%%%%%%%%%

\clearpage

\begin{widetext}

\subsection{\large Supplementary Material}

\section{Fundamental channels and scattering amplitudes} \label{sec-amp}

Here we show that the four fundamental RIXS channels $\eta = \{
0,x,y,z \}$ do not interfere and derive their corresponding
scattering amplitudes $A_{\eta} (m, \mathbf{q})$. First, the
spin-space scattering amplitude $A_{\alpha \beta} (m, \mathbf{q})$
in Eq.~(2) of the main text is rewritten as
\begin{eqnarray}
A_{\alpha \beta} (m, \mathbf{q}) &=& -i \sum_{\mathbf{r}} e^{i
\mathbf{q} \cdot \mathbf{r}} \int_0^{\infty} dt \, e^{-\Gamma t + i
\Omega t} \, \langle m | d_{\mathbf{r}, \alpha}^{\phantom{\dag}}
\left[ \sum_{\tilde{n}_{\mathbf{r}}} e^{-i E_{\tilde{n}} t} \, |
\tilde{n}_{\mathbf{r}} \rangle \langle \tilde{n}_{\mathbf{r}} |
\right] d_{\mathbf{r}, \beta}^{\dag} | 0 \rangle
\nonumber \\
&=& -i \sum_{\mathbf{r}} \int_0^{\infty} dt \, e^{-\Gamma t + i
\Omega t + i \mathbf{q} \cdot \mathbf{r}} \, \Lambda_{\alpha \beta}
(m, t, {\mathbf{r}}), \label{eq-amp-A-1}
\end{eqnarray}
where the kernel of the integral is a local quantum quench of the
form
\begin{eqnarray}
\Lambda_{\alpha \beta} (m, t, {\mathbf{r}}) = \langle m |
d_{\mathbf{r}, \alpha}^{\phantom{\dag}} \, e^{-it \tilde{H}
(\mathbf{r})} \, d_{\mathbf{r}, \beta}^{\dag} | 0 \rangle.
\label{eq-amp-L-1}
\end{eqnarray}
The Hamiltonian $\tilde{H} (\mathbf{r})$ corresponds to the Kitaev
model with a single vacancy at site $\mathbf{r}$. However, instead
of actually removing the spin at the vacancy site $\mathbf{r}$, we
neutralize it by switching off its couplings to its neighbors and
demanding that it is always in the spin-up state. The Hamiltonian
$\tilde{H} (\mathbf{r})$ is then obtained from the Hamiltonian $H$
of the original Kitaev model by switching off the couplings around
site $\mathbf{r}$. Furthermore, the operators $d_{{\mathbf{r}},
\downarrow}^{\dag}$ and $d_{{\mathbf{r}}, \uparrow}^{\dag}$ are
substituted with appropriate projectors given by
\begin{eqnarray}
d_{{\mathbf{r}}, \downarrow}^{\dag} \rightarrow \frac{1}{2} \, (1 +
\sigma_{\mathbf{r}}^z), \qquad d_{{\mathbf{r}}, \uparrow}^{\dag}
\rightarrow \frac{1}{2} \, \sigma_{\mathbf{r}}^x \, (1 -
\sigma_{\mathbf{r}}^z). \label{eq-amp-d}
\end{eqnarray}
From the point of view of the Kitaev model, the creation of a
spin-down (spin-up) electron is equivalent to the annihilation of a
spin-up (spin-down) electron. We therefore first project onto the
subspaces with $\sigma_{\mathbf{r}}^z = \pm 1$ in the two respective
cases. However, in the second case, this projection must also be
followed by an operation $\sigma_{\mathbf{r}}^x$ to ensure that the
vacancy spin at site $\mathbf{r}$ ends up in the
$\sigma_{\mathbf{r}}^z = +1$ state. The amplitude kernels in
Eq.~(\ref{eq-amp-L-1}) then become
\begin{eqnarray}
\Lambda_{\uparrow \uparrow} (m, t, {\mathbf{r}}) &=& \frac{1}{4}
\langle m | (1 - \sigma_{\mathbf{r}}^z) \, \sigma_{\mathbf{r}}^x \,
e^{-it \tilde{H} (\mathbf{r})} \, \sigma_{\mathbf{r}}^x \, (1 -
\sigma_{\mathbf{r}}^z) | 0 \rangle = \frac{1}{2} \langle m | (1 -
\sigma_{\mathbf{r}}^z) \, e^{-it \tilde{H} (\mathbf{r})} | 0
\rangle,
\nonumber \\
\Lambda_{\downarrow \downarrow} (m, t, {\mathbf{r}}) &=& \frac{1}{4}
\langle m | (1 + \sigma_{\mathbf{r}}^z) \, e^{-it \tilde{H}
(\mathbf{r})} \, (1 + \sigma_{\mathbf{r}}^z) | 0 \rangle =
\frac{1}{2} \langle m | (1 + \sigma_{\mathbf{r}}^z) \, e^{-it
\tilde{H} (\mathbf{r})} | 0 \rangle,
\label{eq-amp-L-2} \\
\Lambda_{\uparrow \downarrow} (m, t, {\mathbf{r}}) &=& \frac{1}{4}
\langle m | (1 - \sigma_{\mathbf{r}}^z) \, \sigma_{\mathbf{r}}^x \,
e^{-it \tilde{H} (\mathbf{r})} \, (1 + \sigma_{\mathbf{r}}^z) | 0
\rangle = \frac{1}{2} \langle m | (1 - \sigma_{\mathbf{r}}^z) \,
\sigma_{\mathbf{r}}^x \, e^{-it \tilde{H} (\mathbf{r})} | 0 \rangle,
\nonumber \\
\Lambda_{\downarrow \uparrow} (m, t, {\mathbf{r}}) &=& \frac{1}{4}
\langle m | (1 + \sigma_{\mathbf{r}}^z) \, e^{-it \tilde{H}
(\mathbf{r})} \, \sigma_{\mathbf{r}}^x \, (1 -
\sigma_{\mathbf{r}}^z) | 0 \rangle = \frac{1}{2} \langle m | (1 +
\sigma_{\mathbf{r}}^z) \, \sigma_{\mathbf{r}}^x \, e^{-it \tilde{H}
(\mathbf{r})} | 0 \rangle, \nonumber
\end{eqnarray}
where we can use $[\sigma_{\mathbf{r}}^{\kappa}, \tilde{H}
(\mathbf{r})] = 0$ for all $\kappa = \{ x,y,z \}$ because the
Hamiltonian $\tilde{H} (\mathbf{r})$ does not act on the vacancy
spin. For the four fundamental channels $\eta = \{ 0,x,y,z \}$
introduced in the main text, the relevant scattering amplitudes are
\begin{eqnarray}
A_{\eta} (m, \mathbf{q}) = P_{\eta} \sum_{\alpha, \beta}
\sigma_{\alpha \beta}^{\eta} A_{\alpha \beta} (m, \mathbf{q}) = -i
P_{\eta} \sum_{\mathbf{r}} \int_0^{\infty} dt \, e^{-\Gamma t + i
\Omega t + i \mathbf{q} \cdot \mathbf{r}} \Lambda_{\eta} (m, t,
\mathbf{r}), \label{eq-amp-A-2}
\end{eqnarray}
where the kernels of the integrals are given by
\begin{eqnarray}
\Lambda_{\eta} (m, t, {\mathbf{r}}) = \sum_{\alpha, \beta}
\sigma_{\alpha \beta}^{\eta} \, \Lambda_{\alpha \beta} (m, t,
{\mathbf{r}}). \label{eq-amp-L-3}
\end{eqnarray}
For the four individual channels, these amplitude kernels take the
forms
\begin{eqnarray}
\Lambda_0 (m, t, {\mathbf{r}}) &=& \sum_{\alpha, \beta}
\delta_{\alpha \beta} \Lambda_{\alpha \beta} (m, t, {\mathbf{r}}) =
\Lambda_{\uparrow \uparrow} (m, t, {\mathbf{r}}) +
\Lambda_{\downarrow \downarrow} (m, t, {\mathbf{r}}) = \langle m |
e^{-it \tilde{H} (\mathbf{r})} | 0 \rangle,
\nonumber \\
\Lambda_x (m, t, {\mathbf{r}}) &=& \sum_{\alpha, \beta}
\sigma^x_{\alpha \beta} \Lambda_{\alpha \beta} (m, t, {\mathbf{r}})
= \Lambda_{\uparrow \downarrow} (m, t, {\mathbf{r}}) +
\Lambda_{\downarrow \uparrow} (m, t, {\mathbf{r}}) = \langle m |
\sigma_{\mathbf{r}}^x \, e^{-it \tilde{H} (\mathbf{r})} | 0 \rangle,
\label{eq-amp-L-4} \\
\Lambda_y (m, t, {\mathbf{r}}) &=& \sum_{\alpha, \beta}
\sigma^y_{\alpha \beta} \Lambda_{\alpha \beta} (m, t, {\mathbf{r}})
= -i \Lambda_{\uparrow \downarrow} (m, t, {\mathbf{r}}) + i
\Lambda_{\downarrow \uparrow} (m, t, {\mathbf{r}}) = -\langle m |
\sigma_{\mathbf{r}}^y \, e^{-it \tilde{H} (\mathbf{r})} | 0 \rangle,
\nonumber \\
\Lambda_z (m, t, {\mathbf{r}}) &=& \sum_{\alpha, \beta}
\sigma^z_{\alpha \beta} \Lambda_{\alpha \beta} (m, t, {\mathbf{r}})
= \Lambda_{\uparrow \uparrow} (m, t, {\mathbf{r}}) -
\Lambda_{\downarrow \downarrow} (m, t, {\mathbf{r}}) = -\langle m |
\sigma_{\mathbf{r}}^z \, e^{-it \tilde{H} (\mathbf{r})} | 0 \rangle.
\nonumber
\end{eqnarray}
Since the Hamiltonian $\tilde{H} (\mathbf{r})$ conserves the fluxes,
the possible final states $| m \rangle$ have a definite flux
configuration for each channel $\eta = \{ 0,x,y,z \}$. In
particular, $\Lambda_0 (m, t, {\mathbf{r}})$ can only be non-zero if
$| m \rangle$ has no flux excitations, while $\Lambda_{\kappa} (m,
t, {\mathbf{r}})$ for $\kappa = \{ x,y,z \}$ can only be non-zero if
$| m \rangle$ has two flux excitations separated by a $\kappa$ bond.
The spin-conserving and the three respective non-spin-conserving
channels therefore lead to mutually non-interfering RIXS processes
with additive intensities. From Eqs.~(\ref{eq-amp-A-2}) and
(\ref{eq-amp-L-4}), we immediately recover the corresponding RIXS
amplitudes in Eq.~(3) of the main text.

\section{Spin-conserving RIXS intensity in the few-particle approach} \label{sec-few}

Here we derive an analytic expression for the spin-conserving RIXS
intensity $I_0 (\omega, \mathbf{q})$. According to Eq.~(4) of the
main text, the spin-conserving RIXS amplitude between the ground
state $| 0 \rangle$ and a generic final state $| m \rangle$ is
\begin{equation}
A_0 (m, \mathbf{q}) \propto \sum_{\mathbf{r}} e^{i \mathbf{q} \cdot
\mathbf{r}} \langle m | \bigg[ 1 - \frac{i}{\Gamma} \sum_{\kappa =
x,y,z} J_{\kappa} \sigma_{\mathbf{r}}^{\kappa} \sigma_{\kappa
(\mathbf{r})}^{\kappa} \bigg] | 0 \rangle. \label{eq-few-A-1}
\end{equation}
Since the flux-free configuration of the ground state $| 0 \rangle$
is conserved by the two-operator products
$\sigma_{\mathbf{r}}^{\kappa} \sigma_{\kappa
(\mathbf{r})}^{\kappa}$, these operators can be replaced by their
quadratic-fermion counterparts that correspond to the flux-free
configuration. In terms of the Majorana fermions $c_{A,
\mathbf{r}_A}$ and $c_{B, \mathbf{r}_B}$ introduced into the two
sublattices $A$ and $B$, the RIXS amplitude in
Eq.~(\ref{eq-few-A-1}) then becomes
\begin{equation}
A_0 (m, \mathbf{q}) \propto \sum_{\mathbf{r} \in A} e^{i \mathbf{q}
\cdot \mathbf{r}} \langle m | \bigg[ 1 - \frac{1}{\Gamma}
\sum_{\kappa} J_{\kappa} \, c_{A, \mathbf{r}} \, c_{B, \mathbf{r} +
\hat{\mathbf{r}}_{\kappa}} \bigg] | 0 \rangle + \sum_{\mathbf{r} \in
B} e^{i \mathbf{q} \cdot \mathbf{r}} \langle m | \bigg[ 1 -
\frac{1}{\Gamma} \sum_{\kappa} J_{\kappa} \, c_{A, \mathbf{r} -
\hat{\mathbf{r}}_{\kappa}} c_{B, \mathbf{r}} \bigg] | 0 \rangle.
\label{eq-few-A-2}
\end{equation}
To evaluate this amplitude, we expand the Majorana fermions $c_{A,
\mathbf{r}_A}$ and $c_{B, \mathbf{r}_B}$ in terms of the (complex)
free fermions $\psi_{\mathbf{k}}$ that correspond to the vacuum
state $| 0 \rangle$. These free fermions with momenta $\mathbf{k}$
and energies $\varepsilon_{\mathbf{k}} = 2 |\lambda_{\mathbf{k}}|$
are given by
\begin{equation}
\psi_{\mathbf{k}}^{\phantom{\dag}} = \frac{1} {2 \sqrt{N}}
\sum_{\mathbf{r} \in A} e^{-i \mathbf{k} \cdot \mathbf{r}} \, c_{A,
\mathbf{r}} + \frac{i} {2 \sqrt{N}} \sum_{\mathbf{r} \in B} e^{-i
\mathbf{k} \cdot \mathbf{r} + i \varphi_{\mathbf{k}}} \, c_{B,
\mathbf{r}} \, , \label{eq-few-psi}
\end{equation}
where $N$ is the number of sites in each sublattice, while $e^{i
\varphi_{\mathbf{k}}} \equiv \lambda_{\mathbf{k}} /
|\lambda_{\mathbf{k}}|$ and $\lambda_{\mathbf{k}} \equiv
\sum_{\kappa} J_{\kappa} e^{i \mathbf{k} \cdot
\hat{\mathbf{r}}_{\kappa}}$. Taking the inverse of this basis
transformation, the Majorana fermions in the two sublattices can be
expressed as
\begin{equation}
c_{A, \mathbf{r}} = \frac{1} {\sqrt{N}} \sum_{\mathbf{k}} \left[
e^{i \mathbf{k} \cdot \mathbf{r}} \,
\psi_{\mathbf{k}}^{\phantom{\dag}} + e^{-i \mathbf{k} \cdot
\mathbf{r}} \, \psi_{\mathbf{k}}^{\dag} \right], \qquad c_{B,
\mathbf{r}} = -\frac{i} {\sqrt{N}} \sum_{\mathbf{k}} \left[ e^{i
\mathbf{k} \cdot \mathbf{r} - i \varphi_{\mathbf{k}}} \,
\psi_{\mathbf{k}}^{\phantom{\dag}} - e^{-i \mathbf{k} \cdot
\mathbf{r} + i \varphi_{\mathbf{k}}} \, \psi_{\mathbf{k}}^{\dag}
\right]. \label{eq-few-c}
\end{equation}
Substituting Eq.~(\ref{eq-few-c}) into Eq.~(\ref{eq-few-A-2}), we
find that the only possible final states are $| m \rangle = | 0
\rangle$ with no excitations at all and $| m \rangle \neq | 0
\rangle$ with two fermion excitations. Since we are not interested
in elastic processes corresponding to $| m \rangle = | 0 \rangle$,
we restrict our attention to final states $| m \rangle = |
\mathbf{k}_1, \mathbf{k}_2 \rangle$ with two fermion excitations at
momenta $\mathbf{k}_1$ and $\mathbf{k}_2$. For such a final state
with energy $E_m = \varepsilon_{\mathbf{k}_1} +
\varepsilon_{\mathbf{k}_2}$, the absolute value of the RIXS
amplitude in Eq.~(\ref{eq-few-A-2}) becomes
\begin{eqnarray}
| A_0 (m, \mathbf{q}) | &\propto& \frac{1} {\Gamma N} \Bigg|
\sum_{\mathbf{r} \in A} e^{i \mathbf{q} \cdot \mathbf{r}} \,
\sum_{\kappa} J_{\kappa} \left( e^{-i \mathbf{k}_1 \cdot \mathbf{r}
- i \mathbf{k}_2 \cdot (\mathbf{r} + \hat{\mathbf{r}}_{\kappa}) + i
\varphi_{\mathbf{k}_2}} \langle m | \psi_{\mathbf{k}_1}^{\dag}
\psi_{\mathbf{k}_2}^{\dag} | 0 \rangle + e^{-i \mathbf{k}_2 \cdot
\mathbf{r} - i \mathbf{k}_1 \cdot (\mathbf{r} +
\hat{\mathbf{r}}_{\kappa}) + i \varphi_{\mathbf{k}_1}} \langle m |
\psi_{\mathbf{k}_2}^{\dag} \psi_{\mathbf{k}_1}^{\dag} | 0 \rangle
\right) \nonumber \\
&& + \sum_{\mathbf{r} \in B} e^{i \mathbf{q} \cdot \mathbf{r}} \,
\sum_{\kappa} J_{\kappa} \left( e^{-i \mathbf{k}_1 \cdot (\mathbf{r}
- \hat{\mathbf{r}}_{\kappa}) - i \mathbf{k}_2 \cdot \mathbf{r} + i
\varphi_{\mathbf{k}_2}} \langle m | \psi_{\mathbf{k}_1}^{\dag}
\psi_{\mathbf{k}_2}^{\dag} | 0 \rangle + e^{-i \mathbf{k}_2 \cdot
(\mathbf{r} - \hat{\mathbf{r}}_{\kappa}) - i \mathbf{k}_1 \cdot
\mathbf{r} + i \varphi_{\mathbf{k}_1}} \langle m |
\psi_{\mathbf{k}_2}^{\dag} \psi_{\mathbf{k}_1}^{\dag} | 0 \rangle
\right) \Bigg|
\nonumber \\
&=& \frac{1} {\Gamma N} \Bigg| \sum_{\mathbf{r} \in A} e^{i
(\mathbf{q} - \mathbf{k}_1 - \mathbf{k}_2) \cdot \mathbf{r}} \,
\bigg[ \sum_{\kappa} J_{\kappa} e^{-i \mathbf{k}_2 \cdot
\hat{\mathbf{r}}_{\kappa} + i \varphi_{\mathbf{k}_2}} -
\sum_{\kappa} J_{\kappa} e^{-i \mathbf{k}_1 \cdot
\hat{\mathbf{r}}_{\kappa} + i \varphi_{\mathbf{k}_1}} \bigg]
\label{eq-few-A-3} \\
&& + \sum_{\mathbf{r} \in B} e^{i (\mathbf{q} - \mathbf{k}_1 -
\mathbf{k}_2) \cdot \mathbf{r}} \, \bigg[ \sum_{\kappa} J_{\kappa}
e^{i \mathbf{k}_1 \cdot \hat{\mathbf{r}}_{\kappa} + i
\varphi_{\mathbf{k}_2}} - \sum_{\kappa} J_{\kappa} e^{i \mathbf{k}_2
\cdot \hat{\mathbf{r}}_{\kappa} + i \varphi_{\mathbf{k}_1}} \bigg]
\Bigg| \nonumber \\
&\propto& \delta_{\mathbf{q} - \mathbf{k}_1 - \mathbf{k}_2 -
\mathbf{G}} \left| \left( \varepsilon_{\mathbf{k}_1} -
\varepsilon_{\mathbf{k}_2} \right) \left( 1 - e^{i \mathbf{G} \cdot
\hat{\mathbf{r}}} \, e^{i \varphi_{\mathbf{k}_1}} \, e^{i
\varphi_{\mathbf{k}_2}} \right) \right| \, = \, \delta_{\mathbf{q} -
\mathbf{k}_1 - \mathbf{k}_2 - \mathbf{G}} \left| \left(
\varepsilon_{\mathbf{k}_1} - \varepsilon_{\mathbf{q} - \mathbf{k}_1}
\right) \left( 1 - e^{i \varphi_{\mathbf{k}_1}} \, e^{i
\varphi_{\mathbf{q} - \mathbf{k}_1}} \right) \right|, \nonumber
\end{eqnarray}
where $\mathbf{G}$ is a generic reciprocal lattice vector and
$\hat{\mathbf{r}}$ can be chosen as either one of the three bond
vectors $\hat{\mathbf{r}}_{x,y,z}$. The first factor
$(\varepsilon_{\mathbf{k}_1} - \varepsilon_{\mathbf{q} -
\mathbf{k}_1})$ appears because the fermions
$\psi_{\mathbf{k}_1}^{\dag}$ and $\psi_{\mathbf{k}_2}^{\dag}$ can be
created in two different orders, while the second factor $(1 - e^{i
\varphi_{\mathbf{k}_1}} \, e^{i \varphi_{\mathbf{q} -
\mathbf{k}_1}})$ appears because the fermion creation can occur at
the sites of either sublattice $A$ or sublattice $B$. The
spin-conserving RIXS intensity is then given by
\begin{eqnarray}
I_0 (\omega, \mathbf{q}) &=& \sum_m |A_0 (m, \mathbf{q})|^2 \,
\delta (\omega - E_m) \propto \sum_{\mathbf{k}_1, \mathbf{k}_2}
\delta_{\mathbf{q} - \mathbf{k}_1 - \mathbf{k}_2 - \mathbf{G}}
\left[ \varepsilon_{\mathbf{k}_1} - \varepsilon_{\mathbf{q} -
\mathbf{k}_1} \right]^2 \left| 1 - e^{i \varphi_{\mathbf{k}_1}} \,
e^{i \varphi_{\mathbf{q} - \mathbf{k}_1}} \right|^2 \delta (\omega -
\varepsilon_{\mathbf{k}_1} - \varepsilon_{\mathbf{k}_2})
\nonumber \\
&=& \sum_{\mathbf{k}_1} \left[ \varepsilon_{\mathbf{k}_1} -
\varepsilon_{\mathbf{q} - \mathbf{k}_1} \right]^2 \left| 1 - e^{i
\varphi_{\mathbf{k}_1}} \, e^{i \varphi_{\mathbf{q} - \mathbf{k}_1}}
\right|^2 \delta (\omega - \varepsilon_{\mathbf{k}_1} -
\varepsilon_{\mathbf{q} - \mathbf{k}_1}). \label{eq-few-I}
\end{eqnarray}
Relabeling $\mathbf{k}_1$ into $\mathbf{k}$, and turning the sum in
$\mathbf{k}$ into an integral, we immediately recover the
corresponding RIXS intensity in Eq.~(5) of the main text. Note that
this result does not straightforwardly generalize to the
non-spin-conserving channels because the final state $| m \rangle$
has a different flux configuration and hence different free fermions
with respect to the ground state $| 0 \rangle$.

\section{Determinant approach} \label{sec-det}

\subsection{RIXS intensities as ground-state expectation values} \label{sec-int}

Starting from Eq.~(4) of the main text and recognizing that $1 - i
\tilde{H} (\mathbf{r}) / \Gamma = \exp [- i \tilde{H} (\mathbf{r}) /
\Gamma]$ up to first order in $J_{x,y,z} / \Gamma$, the RIXS
intensity in each fundamental channel $\eta = \{ 0,x,y,z \}$ takes
the form
\begin{eqnarray}
I_{\eta} (\omega, \mathbf{q}) &=& \sum_m |A_{\eta} (m,
\mathbf{q})|^2 \, \delta (\omega - E_m)
\label{eq-int-I-1} \\
&\propto& \sum_m \delta (\omega - E_m) \sum_{\mathbf{r},
\mathbf{r}'} e^{i \mathbf{q} \cdot (\mathbf{r} - \mathbf{r}')} \,
\langle 0 | \sigma_{\mathbf{r}'}^{\eta} \, e^{i \tilde{H}
(\mathbf{r}') / \Gamma} | m \rangle \langle m | e^{-i \tilde{H}
(\mathbf{r}) / \Gamma} \, \sigma_{\mathbf{r}}^{\eta} | 0 \rangle.
\nonumber
\end{eqnarray}
Choosing a reference site $\mathbf{0}$ in sublattice $A$, the
position $\mathbf{r}$ of any site can be expressed as $\mathbf{r} =
\mathbf{R}$ if it is in sublattice $A$ and as $\mathbf{r} =
\mathbf{R} + \hat{\mathbf{r}}$ if it is in sublattice $B$, where
$\mathbf{R}$ is a generic lattice vector and $\hat{\mathbf{r}}$ is a
bond vector pointing from any site in sublattice $A$ to a
neighboring site in sublattice $B$. Since the Kitaev model is
invariant under an overall translation by a lattice vector
$\mathbf{R}$, the RIXS intensity in Eq.~(\ref{eq-int-I-1}) can then
be rewritten as
\begin{eqnarray}
I_{\eta} (\omega, \mathbf{q}) &\propto& N \sum_m \delta (\omega -
E_m) \sum_{\mathbf{R}} e^{i \mathbf{q} \cdot \mathbf{R}} \, \langle
0 | \left[ \sigma_{\mathbf{0}}^{\eta} \, e^{i \tilde{H} (\mathbf{0})
/ \Gamma} + \sigma_{\hat{\mathbf{r}}}^{\eta} \, e^{i \tilde{H}
(\hat{\mathbf{r}}) / \Gamma - i \mathbf{q} \cdot \hat{\mathbf{r}}}
\right] | m \rangle
\nonumber \\
&& \times \langle m | \left[ e^{-i \tilde{H} (\mathbf{R}) / \Gamma}
\, \sigma_{\mathbf{R}}^{\eta} + e^{-i \tilde{H} (\mathbf{R} +
\hat{\mathbf{r}}) / \Gamma + i \mathbf{q} \cdot \hat{\mathbf{r}}} \,
\sigma_{\mathbf{R} + \hat{\mathbf{r}}}^{\eta} \right] | 0 \rangle
\nonumber \\
&=& \frac{N} {2 \pi} \int_{-\infty}^{+\infty} ds \, e^{i \omega s}
\, K_{\eta} (s, \mathbf{q}), \label{eq-int-I-2}
\end{eqnarray}
where $N$ is the number of sites in each sublattice, and the kernel
of the integral is given by
\begin{eqnarray}
K_{\eta} (s, \mathbf{q}) &=& \sum_{\mathbf{R}} e^{i \mathbf{q} \cdot
\mathbf{R}} \, \langle 0 | \left[ \sigma_{\mathbf{0}}^{\eta} \, e^{i
\tilde{H} (\mathbf{0}) / \Gamma} + \sigma_{\hat{\mathbf{r}}}^{\eta}
\, e^{i \tilde{H} (\hat{\mathbf{r}}) / \Gamma - i \mathbf{q} \cdot
\hat{\mathbf{r}}} \right] \left[ \sum_m e^{-i E_m s} \, | m \rangle
\langle m | \right]
\nonumber \\
&& \times \left[ e^{-i \tilde{H} (\mathbf{R}) / \Gamma} \,
\sigma_{\mathbf{R}}^{\eta} + e^{-i \tilde{H} (\mathbf{R} +
\hat{\mathbf{r}}) / \Gamma + i \mathbf{q} \cdot \hat{\mathbf{r}}} \,
\sigma_{\mathbf{R} + \hat{\mathbf{r}}}^{\eta} \right] | 0 \rangle,
\label{eq-int-K-1} \\ \nonumber \\
&=& \sum_{\mathbf{R}} e^{i \mathbf{q} \cdot \mathbf{R}} \, \langle 0
| \left[ \sigma_{\mathbf{0}}^{\eta} \, e^{i \tilde{H} (\mathbf{0}) /
\Gamma} + \sigma_{\hat{\mathbf{r}}}^{\eta} \, e^{i \tilde{H}
(\hat{\mathbf{r}}) / \Gamma - i \mathbf{q} \cdot \hat{\mathbf{r}}}
\right] e^{-i s H} \left[ e^{-i \tilde{H} (\mathbf{R}) / \Gamma} \,
\sigma_{\mathbf{R}}^{\eta} + e^{-i \tilde{H} (\mathbf{R} +
\hat{\mathbf{r}}) / \Gamma + i \mathbf{q} \cdot \hat{\mathbf{r}}} \,
\sigma_{\mathbf{R} + \hat{\mathbf{r}}}^{\eta} \right] | 0 \rangle.
\nonumber
\end{eqnarray}
Since the sum in $\mathbf{R}$ runs over $N$ lattice points, the
kernels for a given value of $s$ can be obtained by evaluating $4N$
ground-state expectation values of the Kitaev model. For the
spin-conserving channel, all $4N$ expectation values are non-zero in
general. However, for the non-spin-conserving channels, only $4$
expectation values are non-zero. In particular, for $\eta = \kappa =
\{ x,y,z \}$, if we set $\hat{\mathbf{r}} =
\hat{\mathbf{r}}_{\kappa}$, such that $\mathbf{0}$ and
$\hat{\mathbf{r}}_{\kappa}$ are neighboring sites connected by a
$\kappa$ bond, the expectation values with $\mathbf{R} \neq
\mathbf{0}$ all vanish because there is a mismatch between the
original and the final flux configurations. The integral kernel in
Eq.~(\ref{eq-int-K-1}) then becomes
\begin{equation}
K_{\kappa} (s, \mathbf{q}) = \langle 0 | \left[
\sigma_{\mathbf{0}}^{\kappa} \, e^{i \tilde{H} (\mathbf{0}) /
\Gamma} + \sigma_{\hat{\mathbf{r}}_{\kappa}}^{\kappa} \, e^{i
\tilde{H} (\hat{\mathbf{r}}_{\kappa}) / \Gamma - i \mathbf{q} \cdot
\hat{\mathbf{r}}_{\kappa}} \right] e^{-i s H} \left[ e^{-i \tilde{H}
(\mathbf{0}) / \Gamma} \, \sigma_{\mathbf{0}}^{\kappa} + e^{-i
\tilde{H} (\hat{\mathbf{r}}_{\kappa}) / \Gamma + i \mathbf{q} \cdot
\hat{\mathbf{r}}_{\kappa}} \,
\sigma_{\hat{\mathbf{r}}_{\kappa}}^{\kappa} \right] | 0 \rangle.
\label{eq-int-K-2}
\end{equation}
This form of the kernel is identical to the one presented in the
main text. In the following, we restrict our current approach to the
non-spin-conserving channels and the corresponding form in
Eq.~(\ref{eq-int-K-2}) because the general form in
Eq.~(\ref{eq-int-K-1}) requires $N$ times more computational effort
and is therefore not practical to use.

\subsection{Reduction to free-fermion expectation values} \label{sec-ferm}

Following its exact solution, the ground-state expectation values of
the Kitaev model can be reduced to those of appropriate free-fermion
problems. The two essentially distinct expectation values in
Eq.~(\ref{eq-int-K-2}) are
\begin{eqnarray}
S_{\kappa}^{(1)} &=& \langle 0 | \sigma_{\mathbf{0}}^{\kappa} \,
e^{i \tilde{H} (\mathbf{0}) / \Gamma} \, e^{-i s H} \, e^{-i
\tilde{H} (\mathbf{0}) / \Gamma} \, \sigma_{\mathbf{0}}^{\kappa} | 0
\rangle,
\label{eq-ferm-S-1} \\
S_{\kappa}^{(2)} &=& \langle 0 |
\sigma_{\hat{\mathbf{r}}_{\kappa}}^{\kappa} \, e^{i \tilde{H}
(\hat{\mathbf{r}}_{\kappa}) / \Gamma} \, e^{-i s H} \, e^{-i
\tilde{H} (\mathbf{0}) / \Gamma} \, \sigma_{\mathbf{0}}^{\kappa} | 0
\rangle. \nonumber
\end{eqnarray}
In each expectation value, we first exchange the operator
$\sigma_{\mathbf{0}}^{\kappa}$ on the right side with the three
exponential operators. Taking care of the appropriate spin
commutation relations, the expectation values then become
\begin{eqnarray}
S_{\kappa}^{(1)} &=& \langle 0 | \sigma_{\mathbf{0}}^{\kappa}
\sigma_{\mathbf{0}}^{\kappa} \, e^{i \tilde{H} (\mathbf{0}) /
\Gamma} \, e^{-is \hat{H}_{\kappa} (\mathbf{0})} \, e^{-i \tilde{H}
(\mathbf{0}) / \Gamma} | 0 \rangle = \langle 0 | e^{i \tilde{H}
(\mathbf{0}) / \Gamma} \, e^{-is \hat{H}_{\kappa} (\mathbf{0})} \,
e^{-i \tilde{H} (\mathbf{0}) / \Gamma} | 0 \rangle,
\nonumber \\
S_{\kappa}^{(2)} &=& \langle 0 | \sigma_{\mathbf{0}}^{\kappa}
\sigma_{\hat{\mathbf{r}}_{\kappa}}^{\kappa} \, e^{i
\check{H}_{\kappa} (\mathbf{0}, \hat{\mathbf{r}}_{\kappa}) / \Gamma}
\, e^{-is \hat{H}_{\kappa} (\mathbf{0})} \, e^{-i \tilde{H}
(\mathbf{0}) / \Gamma} | 0 \rangle, \label{eq-ferm-S-2}
\end{eqnarray}
where the Hamiltonians $\hat{H}_{\kappa} (\mathbf{0})$ and
$\check{H}_{\kappa} (\mathbf{0}, \hat{\mathbf{r}}_{\kappa})$ are
obtained from the Hamiltonians $H$ and $\tilde{H}
(\hat{\mathbf{r}}_{\kappa})$, respectively, by reversing the
couplings of the non-$\kappa$ bonds around site $\mathbf{0}$. Since
the flux-free configuration of the ground state $| 0 \rangle$ is
conserved by the two-operator product $\sigma_{\mathbf{0}}^{\kappa}
\sigma_{\hat{\mathbf{r}}_{\kappa}}^{\kappa}$ as well as the
exponential operators inside both expectation values, these
operators can all be replaced by their quadratic-fermion
counterparts that correspond to the flux-free configuration. In
terms of the Majorana fermions $c_{A, \mathbf{r}_A}$ and $c_{B,
\mathbf{r}_B}$ introduced into the two sublattices $A$ and $B$, the
resulting substitutions are
\begin{equation}
\sigma_{\mathbf{0}}^{\kappa}
\sigma_{\hat{\mathbf{r}}_{\kappa}}^{\kappa} \rightarrow -i c_{A,
\mathbf{0}} \, c_{B, \hat{\mathbf{r}}_{\kappa}}, \qquad H
\rightarrow \sum_{\mathbf{r}_A, \mathbf{r}_B} i M_{\mathbf{r}_A,
\mathbf{r}_B} c_{A, \mathbf{r}_A} c_{B, \mathbf{r}_B},
\label{eq-ferm-sub}
\end{equation}
where the coupling matrix $M_{\mathbf{r}_A, \mathbf{r}_B}$ specifies
the coupling between sites $\mathbf{r}_A$ and $\mathbf{r}_B$. In
particular, $M_{\mathbf{r}_A, \mathbf{r}_B} = J_{\kappa}$ if
$\mathbf{r}_A$ and $\mathbf{r}_B$ are neighboring sites connected by
a $\kappa$ bond and $M_{\mathbf{r}_A, \mathbf{r}_B} = 0$ if they are
not neighboring sites. The substitutions for the Hamiltonians
$\tilde{H} (\mathbf{0})$, $\hat{H}_{\kappa} (\mathbf{0})$, and
$\check{H}_{\kappa} (\mathbf{0}, \hat{\mathbf{r}}_{\kappa})$ are
analogous to that for the Hamiltonian $H$ with respective coupling
matrices $\tilde{M}_{\mathbf{r}_A, \mathbf{r}_B} (\mathbf{0})$,
$\hat{M}_{\mathbf{r}_A, \mathbf{r}_B}^{\kappa} (\mathbf{0})$, and
$\check{M}_{\mathbf{r}_A, \mathbf{r}_B}^{\kappa} (\mathbf{0},
\hat{\mathbf{r}}_{\kappa})$. These coupling matrices are related to
the original coupling matrix by
\begin{eqnarray}
\tilde{M}_{\mathbf{r}_A, \mathbf{r}_B} (\mathbf{0}) &=&
M_{\mathbf{r}_A, \mathbf{r}_B} (1 - \delta_{\mathbf{r}_A,
\mathbf{0}}),
\nonumber \\
\hat{M}_{\mathbf{r}_A, \mathbf{r}_B}^{\kappa} (\mathbf{0}) &=&
M_{\mathbf{r}_A, \mathbf{r}_B} (1 - 2 \delta_{\mathbf{r}_A,
\mathbf{0}} + 2 \delta_{\mathbf{r}_B, \hat{\mathbf{r}}_{\kappa}}),
\label{eq-ferm-M} \\
\check{M}_{\mathbf{r}_A, \mathbf{r}_B}^{\kappa} (\mathbf{0},
\hat{\mathbf{r}}_{\kappa}) &=& M_{\mathbf{r}_A, \mathbf{r}_B} (1 - 2
\delta_{\mathbf{r}_A, \mathbf{0}}) (1 - \delta_{\mathbf{r}_B,
\hat{\mathbf{r}}_{\kappa}}). \nonumber
\end{eqnarray}
Using the substitutions in Eq.~(\ref{eq-ferm-sub}), the ground-state
expectation values in Eq.~(\ref{eq-ferm-S-2}) then become
\begin{eqnarray}
S_{\kappa}^{(1)} &=& \langle 0 | \exp \left[ -\frac{1} {\Gamma}
\sum_{\mathbf{r}_A, \mathbf{r}_B} \tilde{M}_{\mathbf{r}_A,
\mathbf{r}_B} (\mathbf{0}) \, c_{A, \mathbf{r}_A} c_{B,
\mathbf{r}_B} \right] \exp \left[ \sum_{\mathbf{r}_A, \mathbf{r}_B}
s \, \hat{M}_{\mathbf{r}_A, \mathbf{r}_B}^{\kappa} (\mathbf{0}) \,
c_{A, \mathbf{r}_A} c_{B, \mathbf{r}_B} \right]
\nonumber \\
&& \times \exp \left[ \frac{1} {\Gamma} \sum_{\mathbf{r}_A,
\mathbf{r}_B} \tilde{M}_{\mathbf{r}_A, \mathbf{r}_B} (\mathbf{0}) \,
c_{A, \mathbf{r}_A} c_{B, \mathbf{r}_B} \right] | 0 \rangle,
\label{eq-ferm-S-3} \\
S_{\kappa}^{(2)} &=& \langle 0 | \left( -i c_{A, \mathbf{0}} \,
c_{B, \hat{\mathbf{r}}_{\kappa}} \right) \exp \left[ -\frac{1}
{\Gamma} \sum_{\mathbf{r}_A, \mathbf{r}_B} \check{M}_{\mathbf{r}_A,
\mathbf{r}_B}^{\kappa} (\mathbf{0}, \hat{\mathbf{r}}_{\kappa}) \,
c_{A, \mathbf{r}_A} c_{B, \mathbf{r}_B} \right]
\nonumber \\
&& \times \exp \left[ \sum_{\mathbf{r}_A, \mathbf{r}_B} s \,
\hat{M}_{\mathbf{r}_A, \mathbf{r}_B}^{\kappa} (\mathbf{0}) \, c_{A,
\mathbf{r}_A} c_{B, \mathbf{r}_B} \right] \exp \left[ \frac{1}
{\Gamma} \sum_{\mathbf{r}_A, \mathbf{r}_B} \tilde{M}_{\mathbf{r}_A,
\mathbf{r}_B} (\mathbf{0}) \, c_{A, \mathbf{r}_A} c_{B,
\mathbf{r}_B} \right] | 0 \rangle. \nonumber
\end{eqnarray}
Note that the three exponential operators in each expectation value
are not straightforward to combine into a single exponential
operator because the quadratic operators in their arguments do not
commute with one another.

\subsection{Functional determinants via fermion doubling} \label{sec-doub}

Due to their free-fermion forms, the expectation values in
Eq.~(\ref{eq-ferm-S-3}) can be evaluated as functional determinants
for any finite-size system. Since they are given in terms of
Majorana fermions, it is natural to employ fermion doubling and turn
their quadratic Majorana fermion terms into quadratic
number-conserving complex fermion terms. First, we define Majorana
fermion copies $c_{A, \mathbf{r}_A}'$ and $c_{B, \mathbf{r}_B}'$ of
the original Majorana fermions $c_{A, \mathbf{r}_A}$ and $c_{B,
\mathbf{r}_B}$, and introduce corresponding complex fermions given
by
\begin{equation}
f_{A, \mathbf{r}_A} = \frac{1}{2} \left( c_{A, \mathbf{r}_A} + i
c_{A, \mathbf{r}_A}' \right), \qquad f_{B, \mathbf{r}_B} =
\frac{i}{2} \left( c_{B, \mathbf{r}_B} + i c_{B, \mathbf{r}_B}'
\right). \label{eq-doub-f}
\end{equation}
For the original Hamiltonian $H$, the doubled fermion Hamiltonian
then reads
\begin{eqnarray}
H + H' &\rightarrow& \sum_{\mathbf{r}_A, \mathbf{r}_B} \left( i
M_{\mathbf{r}_A, \mathbf{r}_B} c_{A, \mathbf{r}_A} c_{B,
\mathbf{r}_B} + i M_{\mathbf{r}_A, \mathbf{r}_B} c_{A,
\mathbf{r}_A}' c_{B, \mathbf{r}_B}' \right) = \sum_{\mathbf{r}_A,
\mathbf{r}_B} 2 M_{\mathbf{r}_A, \mathbf{r}_B} \left( f_{A,
\mathbf{r}_A}^{\dag} f_{B, \mathbf{r}_B}^{\phantom{\dag}} + f_{B,
\mathbf{r}_B}^{\dag} f_{A, \mathbf{r}_A}^{\phantom{\dag}} \right)
\nonumber \\
&\equiv& \left( \begin{array}{cc} f_A^{\dag} & f_B^{\dag}
\end{array} \right) \cdot \left( \begin{array}{cc} 0 & 2M \\ 2M^T
& 0 \end{array} \right) \cdot \left( \begin{array}{cc}
f_A^{\phantom{\dag}} \\ f_B^{\phantom{\dag}} \end{array} \right)
\equiv f^{\dag} \cdot \mathcal{H} \cdot f. \label{eq-doub-H-1}
\end{eqnarray}
This quadratic Hamiltonian conserves the total number of complex
fermions. Furthermore, via the singular-value decomposition $2M =
U^T \cdot \varepsilon \cdot V$, it can be rewritten in the
free-fermion form
\begin{equation}
H + H' \rightarrow \sum_{\mathbf{k}} \varepsilon_{\mathbf{k}} \left(
\phi_{\mathbf{k},+}^{\dag} \phi_{\mathbf{k},+}^{\phantom{\dag}} -
\phi_{\mathbf{k},-}^{\dag} \phi_{\mathbf{k},-}^{\phantom{\dag}}
\right) \equiv \left( \begin{array}{cc} \phi_{+}^{\dag} &
\phi_{-}^{\dag} \end{array} \right) \cdot \left( \begin{array}{cc}
\varepsilon & 0 \\ 0 & -\varepsilon \end{array} \right) \cdot \left(
\begin{array}{cc} \phi_{+}^{\phantom{\dag}} \\
\phi_{-}^{\phantom{\dag}} \end{array} \right), \label{eq-doub-H-2}
\end{equation}
where the fermion energies $\varepsilon_{\mathbf{k}}$ at the
respective momenta $\mathbf{k}$ are the (non-negative) elements of
the diagonal matrix $\varepsilon$, and the free fermions themselves
are given in terms of the orthogonal matrices $U$ and $V$ by
\begin{equation}
\phi \equiv \left( \begin{array}{c} \phi_{+}^{\phantom{\dag}} \\
\phi_{-}^{\phantom{\dag}} \end{array} \right) = \frac{1}{\sqrt{2}}
\left( \begin{array}{cc} U & V \\ U & -V \end{array} \right) \cdot
\left( \begin{array}{c} f_A^{\phantom{\dag}} \\
f_B^{\phantom{\dag}} \end{array} \right) \equiv \mathcal{W} \cdot f.
\label{eq-doub-phi}
\end{equation}
Importantly, the ground state $| 0 \rangle \otimes | 0' \rangle$ of
the doubled model is not the vacuum state $| \Phi \rangle$ of the
free fermions $\phi_{\mathbf{k}, \pm}$ because the free fermions
$\phi_{\mathbf{k},-}$ have negative energies in
Eq.~(\ref{eq-doub-H-2}). Instead, the doubled ground state is
\begin{equation}
| 0 \rangle \otimes | 0' \rangle = \prod_{\mathbf{k}}
\phi_{\mathbf{k},-}^{\dag} \, | \Phi \rangle. \label{eq-doub-gs}
\end{equation}
For the remaining Hamiltonians $\tilde{H} (\mathbf{0})$,
$\hat{H}_{\kappa} (\mathbf{0})$, and $\check{H}_{\kappa}
(\mathbf{0}, \hat{\mathbf{r}}_{\kappa})$, the doubled Hamiltonians
and the doubled ground states can be expressed in the same way.
Using this doubling procedure, the square of the expectation value
$S_{\kappa}^{(1)}$ in Eq.~(\ref{eq-ferm-S-3}) becomes
\begin{eqnarray}
\left[ S_{\kappa}^{(1)} \right]^2 &=& \Big( \langle 0 | \otimes
\langle 0' | \Big) \exp \left[ \frac{i} {\Gamma} \left\{ \tilde{H}
(\mathbf{0}) + \tilde{H}' (\mathbf{0}) \right\} \right] \exp \left[
-is \left( \hat{H}_{\kappa} (\mathbf{0}) + \hat{H}_{\kappa}'
(\mathbf{0}) \right) \right] \exp \left[ -\frac{i} {\Gamma} \left\{
\tilde{H} (\mathbf{0}) + \tilde{H}' (\mathbf{0}) \right\} \right]
\Big( | 0 \rangle \otimes | 0' \rangle \Big)
\nonumber \\
&=& \langle \Phi | \Big( \prod_{\mathbf{k}}
\phi_{\mathbf{k},-}^{\phantom{\dag}} \Big) \exp \left[ f^{\dag}
\cdot \left\{ i \tilde{\mathcal{H}} (\mathbf{0}) / \Gamma \right\}
\cdot f \right] \exp \left[ f^{\dag} \cdot \left( -is
\hat{\mathcal{H}}_{\kappa} (\mathbf{0}) \right) \cdot f \right] \exp
\left[ f^{\dag} \cdot \left\{ -i \tilde{\mathcal{H}} (\mathbf{0}) /
\Gamma \right\} \cdot f \right] \Big( \prod_{\mathbf{k}}
\phi_{\mathbf{k},-}^{\dag} \Big) | \Phi \rangle.
\nonumber \\
\label{eq-doub-S-1}
\end{eqnarray}
Furthermore, due to the relations $-i c_{A,
\mathbf{0}}^{\phantom{'}} \, c_{A, \mathbf{0}}' = \exp [i \pi f_{A,
\mathbf{0}}^{\dag} \, f_{A, \mathbf{0}}^{\phantom{\dag}}]$ and $-i
c_{B, \mathbf{\hat{r}}_{\kappa}}^{\phantom{'}} c_{B,
\mathbf{\hat{r}}_{\kappa}}' = \exp [i \pi f_{B,
\mathbf{\hat{r}}_{\kappa}}^{\dag} f_{B,
\mathbf{\hat{r}}_{\kappa}}^{\phantom{\dag}}]$, the square of the
expectation value $S_{\kappa}^{(2)}$ in Eq.~(\ref{eq-ferm-S-3})
takes the form
\begin{eqnarray}
\left[ S_{\kappa}^{(2)} \right]^2 &=& -\langle \Phi | \Big(
\prod_{\mathbf{k}} \phi_{\mathbf{k},-}^{\phantom{\dag}} \Big) \exp
\left[ f^{\dag} \cdot \left\{ i \pi \mathcal{F} (\mathbf{0})
\right\} \cdot f \right] \exp \left[ f^{\dag} \cdot \left\{ i \pi
\mathcal{F} (\mathbf{\mathbf{\hat{r}}_{\kappa}}) \right\} \cdot f
\right] \exp \left[ f^{\dag} \cdot \Big\{ i
\check{\mathcal{H}}_{\kappa} (\mathbf{0}, \mathbf{\hat{r}}_{\kappa})
/ \Gamma \Big\} \cdot f \right]
\nonumber \\
&& \times \exp \left[ f^{\dag} \cdot \left\{ -is
\hat{\mathcal{H}}_{\kappa} (\mathbf{0}) \right\} \cdot f \right]
\exp \left[ f^{\dag} \cdot \left\{ -i \tilde{\mathcal{H}}
(\mathbf{0}) / \Gamma \right\} \cdot f \right] \Big(
\prod_{\mathbf{k}} \phi_{\mathbf{k},-}^{\dag} \Big) | \Phi \rangle,
\label{eq-doub-S-2}
\end{eqnarray}
where the matrix elements of $\mathcal{F} (\mathbf{r})$ are given by
$\mathcal{F}_{\mathbf{r}_1, \mathbf{r}_2} (\mathbf{r}) =
\delta_{\mathbf{r}, \mathbf{r}_1} \delta_{\mathbf{r},
\mathbf{r}_2}$. Note that quadratic terms of the Majorana fermion
copies always commute with quadratic terms of the original Majorana
fermions. The doubled expectation values in Eqs.~(\ref{eq-doub-S-1})
and (\ref{eq-doub-S-2}) are then evaluated via a special case of the
Baker$-$Hausdorff lemma:
\begin{equation}
\exp \left[ f^{\dag} \cdot \Theta \cdot f \right] \left( \lambda
\cdot f^{\dag} \right) \exp \left[ -f^{\dag} \cdot \Theta \cdot f
\right] = \left( e^{\Theta} \cdot \lambda \right) \cdot f^{\dag}.
\label{eq-doub-sc-lemma}
\end{equation}
In particular, this formula shows that exchanging each exponential
operator with the creation operators at the right side of the
expectation value is equivalent to a basis transformation of the
doubled fermions. Recalling that $| \Phi \rangle$ is the vacuum
state of these fermions, the doubled expectation values in
Eqs.~(\ref{eq-doub-S-1}) and (\ref{eq-doub-S-2}) become
\begin{eqnarray}
\left[ S_{\kappa}^{(1)} \right]^2 &=& \langle \Phi | \Big(
\prod_{\mathbf{k}} \phi_{\mathbf{k},-}^{\phantom{\dag}} \Big) \Big(
\prod_{\mathbf{k}} \Big[ \sum_{\mathbf{k}'} \left\{
\mathcal{S}_{\mathbf{k'}, +, \mathbf{k}, -}^{(1)} \,
\phi_{\mathbf{k}',+}^{\dag} + \mathcal{S}_{\mathbf{k'}, -,
\mathbf{k}, -}^{(1)} \, \phi_{\mathbf{k}',-}^{\dag} \right\} \Big]
\Big) | \Phi \rangle = \det \left[ \mathcal{S}_{-,-}^{(1)} \right],
\label{eq-doub-S-3} \\
\left[ S_{\kappa}^{(2)} \right]^2 &=& -\langle \Phi | \Big(
\prod_{\mathbf{k}} \phi_{\mathbf{k},-}^{\phantom{\dag}} \Big) \Big(
\prod_{\mathbf{k}} \Big[ \sum_{\mathbf{k}'} \left\{
\mathcal{S}_{\mathbf{k'}, +, \mathbf{k}, -}^{(2)} \,
\phi_{\mathbf{k}',+}^{\dag} + \mathcal{S}_{\mathbf{k'}, -,
\mathbf{k}, -}^{(2)} \, \phi_{\mathbf{k}',-}^{\dag} \right\} \Big]
\Big) | \Phi \rangle = -\det \left[ \mathcal{S}_{-,-}^{(2)} \right],
\nonumber
\end{eqnarray}
where the unitary basis transformation matrices $\mathcal{S}^{(1)}$
and $\mathcal{S}^{(2)}$ are given by
\begin{eqnarray}
\mathcal{S}^{(1)} = \left( \begin{array}{cc} \mathcal{S}_{+,+}^{(1)}
& \mathcal{S}_{+,-}^{(1)} \\ \mathcal{S}_{-,+}^{(1)} &
\mathcal{S}_{-,-}^{(1)} \end{array} \right) &=& \mathcal{W} \cdot
e^{i \tilde{\mathcal{H}} (\mathbf{0}) / \Gamma} \cdot e^{-is
\hat{\mathcal{H}}_{\kappa} (\mathbf{0})} \cdot e^{-i
\tilde{\mathcal{H}} (\mathbf{0}) / \Gamma} \cdot \mathcal{W}^T,
\label{eq-doub-S-4} \\
\mathcal{S}^{(2)} = \left( \begin{array}{cc} \mathcal{S}_{+,+}^{(2)}
& \mathcal{S}_{+,-}^{(2)} \\ \mathcal{S}_{-,+}^{(2)} &
\mathcal{S}_{-,-}^{(2)} \end{array} \right) &=& \mathcal{W} \cdot
e^{i \pi \mathcal{F} (\mathbf{0})} \cdot e^{i \pi \mathcal{F}
(\mathbf{\hat{r}}_{\kappa})} \cdot e^{i \check{\mathcal{H}}_{\kappa}
(\mathbf{0}, \mathbf{\hat{r}}_{\kappa}) / \Gamma} \cdot e^{-is
\hat{\mathcal{H}}_{\kappa} (\mathbf{0})} \cdot e^{-i
\tilde{\mathcal{H}} (\mathbf{0}) / \Gamma} \cdot \mathcal{W}^T.
\nonumber
\end{eqnarray}
Note that the blocks $\mathcal{S}_{-,-}^{(1,2)}$ of the matrices
$\mathcal{S}^{(1,2)}$ are not unitary in general. To evaluate the
expectation values $S_{\kappa}^{(1)}$ and $S_{\kappa}^{(2)}$
themselves, we must take the square roots of the results in
Eq.~(\ref{eq-doub-S-3}). The corresponding sign ambiguity is fixed
by determining the complex phase of each expectation value at $s =
0$ and demanding that it is a continuous function of $s$.

\clearpage

\end{widetext}

%%%%%%%%%%%%%%%%%%%%%%%%%%%%%%%%%%%%%%%%%%%%%%%%%


\begin{references}
\bibitem{EBZ} Special points with and without dashes are related by
inversion and have identical responses for all $J_{x,y,z}$, while
those with different $x,y,z$ subscripts are related by threefold
rotation and only have identical responses when $J_x = J_y = J_z$.
\end{references}
\end{document}